\documentclass[pra,aps,epsfig,amsmath,twocolumn,amssymb,showpacs]{revtex4} 

\newcommand\be{\begin{eqnarray}}
\newcommand\ee{\end{eqnarray}}
\newcommand\nn{\nonumber}
\usepackage{graphicx} 
\usepackage{amssymb}
\usepackage{amsmath}
\usepackage{amsfonts}

\begin{document}
\title{Decoherence in collective quantum memories for photons}
\author{Claudia Mewes and Michael Fleischhauer}
\affiliation{Fachbereich Physik der Technischen 
Universit\"{a}t Kaiserslautern, D-67663\\
Kaiserslautern, Germany}
\date{\today }
 
\begin{abstract}
The influence of decoherence on the fidelity of  quantum memories
for photonic qubits based on dark-state polaritons in 
atomic ensembles is discussed. It is shown that
despite the large entanglement of the collective storage states 
corresponding to single photons or nonclassical states of light the sensitivity
to decoherence does not scale with the number of atoms. This is due to
the existence of equivalence classes of storage states corresponding to
states with the same number of dark-state polariton excitations but arbitrary
excitations in other polariton modes. Several decoherence processes are
discussed in detail: single-atom spin-flips and dephasing, atom loss and
motion of atoms. 
\end{abstract}
\maketitle

\section{Introduction}
 
One of the essential ingredients for quantum information processing 
with photons  as information carrier \cite{deVinzenco,Zoller-Varenna} 
is a reliable quantum memory
capable of a faithful storage of the quantum state of photons.
They play a key role
in network quantum computing \cite{Mabuchi-PRL-1997}, in long-distance,
secure quantum communication 
and quantum teleportation 
\cite{Duan-PRL-2000,Duan-Nature-2001,Julsgaard-Nature-2001,Kuzmich-Nature-2003,van-der-Wal-Science-2003}. The application to teleportation 
is of particular interest because of its potentials for quantum information
processing with linear optical elements 
\cite{Gottesman-Nature-1999,KLM-Nature-2001}. 
While photons are one of the most easy to handle information carriers,
atoms or similar systems like quantum dots are reliable and long lived
storage units. Furthermore Raman transitions provide a
controllable and decoherence insensitive way of coupling
between light and atoms. The conceptually simplest and for
processing purposes best suited storage system for photonic qubits
are individual atoms. Here coherent transfer techniques have been
developed that allow a controlled transfer of quantum information
from light to the atom and vice versa \cite{Mabuchi-PRL-1997}.
However cavity QED settings in the strong-coupling regime
are required to achieve reasonable fidelities for the state transfer 
\cite{Briegel-LectNotes-1999}. On the other hand if atomic ensembles
are used rather than individual atoms no such requirements exists
and coherent and reversible transfer techniques for individual
photon wavepackets 
\cite{Grobe-PRL-1997,Lukin-PRL-2000,Fleischhauer-PRL-2000,Phillips-PRL-2001,Liu-Nature-2001,Fleischhauer-PRA-2002,Lukin-RMP-2003} and cw light fields
\cite{Kuzmich-PRL-1997,Hald-PRL-1999,Kuzmich-PRL-2000,Schori-PRL-2002}
have been proposed and in part experimentally implemented.

The substantially alleviated requirements for the light-matter
interface in the case of atomic ensembles are due to the
enhanced coupling between collective many-atom states and the
radiation field. The corresponding collective excitations of the
ensemble are highly entangled 
many-particle states if nonclassical states of light are stored. 
So while classical information encoded, 
e.g. in single-particle Raman coherences, can be rather robust against 
decoherence processes this is not a priori clear for 
quantum correlations stored in the ensemble. In fact one might
naivly expect that the livetime of quantum correlations
decreases with the number of atoms involved in the storage state
in which case the system would be practically useless as a 
quantum memory. We therefore analyze in the present paper the 
influence of various decoherence mechanisms on the fidelity of the 
quantum memory. We show that each quantum state of the  radiation
field stored in the atomic ensemble corresponds to a whole
class of many-particle states. It is due to the existence of these
equivalence classes, which represent all states with the same number
of excitations in specific quasi-particle modes, the dark-state polariton, 
and arbitrary
excitations in other modes, that the quantum memory does not
show an enhanced sensitivity to decoherence as compared to single-particle
storage units.

In order to simplify the
discussion we will here restrict ourselves to a quantum memory for a 
single-mode radiation field, realized for example in a weak-coupling
resonator \cite{Lukin-PRL-2000}. In doing so we 
do not need to take into account
decoherence effects on the longitudinal profile
of a stored pulse arizing from atomic motion, which are however important in
free-space configurations \cite{Fleischhauer-PRL-2000,Duan-PRA-2002}. 
First we reexamine the adiabatic transfer scheme of
\cite{Lukin-PRL-2000,Fleischhauer-PRL-2000}
for quantum states of photon wavepackets to 
atomic ensembles in terms of quasi-particles (dark- and bright polaritons)
in Sect.~II. We will show that only specific quasi-particle modes 
are relevant for the storage in the adiabatic limit. 
In Sec.~III we will discuss the effect of different decoherence mechanisms,
individual random spin flips, dephasing of Raman coherences, loss of atoms,
atomic motion and imperfect preparation. It will be shown that
the decoherence rate of the stored quantum state 
does not depend on the number of atoms in all
of these cases. This is because excitations of any quasi-particle mode
other than the relevant dark-polariton mode do not matter in the adiabatic
limit. They do matter, however, if non-adiabatic couplings are taken into
account. We will therefore discuss the effect of decoherence in the
presence of non-adiabatic couplings in Sec.~VI.


\section{Dark- and bright-polaritons, equivalence classes of storage states}
\label{SECTIONII}

\begin{figure}[htb]
\includegraphics[width=5cm]{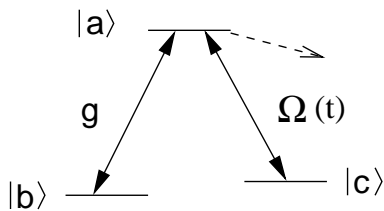} 
\caption{3-level atoms coupled to single quantized resonator mode and classical
control field of (real) Rabi-frequency $\Omega(t)$; $g$-vacuum Rabi-frequency;
the dashed line indicates spontaneous decay.}
\label{3-level}
\end{figure}

Let us consider an ensemble of $N$ 3-level atoms with internal states
$|a\rangle$, $|b\rangle$ and $|c\rangle$ resonantly coupled to a
single quantized mode of a resonator field with mode function 
${\rm e}^{ik_0 z}$ and a classical control field
of Rabi-frequency $\Omega$ and mode function 
${\rm e}^{i \mathbf{k}_1\cdot\mathbf{r}}$ 
as shown in Fig.~\ref{3-level}.
The dynamics of this system is described by a non-hermitian
Hamiltonian ($E_b=\hbar\omega_b=0$):
\be
&&H =\hbar\omega a^\dagger a
+\hbar (\omega_a-i\gamma)\sum_{j=1}^N \sigma_{aa}^j +\hbar\omega_c 
\sum_{j=1}^N \sigma_{cc}^j 
+\label{ham}\\
&&\quad+
 \hbar g \sum_{j = 1}^N   a\sigma_{ab}^j\ {\rm e}^{ik_0 z_j} + 
\hbar\Omega(t) {\rm e}^{-i\nu t} 
\sum_{j = 1}^N  {\rm e}^{i\mathbf{k}_1\cdot\mathbf{r}_j}
\sigma_{ac}^i + {\rm h.c.} .  \nn
\ee
Here $\sigma_{\mu\nu}^j = |\mu\rangle_{jj}\langle \nu|$ is the 
flip operator of the $i$th atom and the vacuum Rabi-frequency
is assumed to be equal for all atoms. For the time being we disregard
atomic motion and thus the phase factors ${\rm  e}^{ik_0 z_j}$ as well
as ${\rm e}^{i\mathbf{k}_1\cdot\mathbf{r}_j}$ will be absorbed into the
definition of the atomic states $|a\rangle_j$ and $|c\rangle_j$. 
We will however come back to the issue of atomic motion in section III.
We also have introduced an imaginary
part to the Hamiltonian to take into account losses from the 
excited state, e.g. via spontaneous emission. 
The complex Hamiltonian emerges from a Lindblad Liouville
operator that includes the decay from the excited state to other 
internal states of the atoms after projection onto the subspace $\{|a\rangle,
|b\rangle,|c\rangle\}$. The model does not take into account relaxation
from the excited state back into the lower levels. Spontaneous emission
into the resonator mode accompanied by a transition from $|a\rangle$ to
$|b\rangle$ is of course automatically included in the model.


\begin{figure}[ht]
\includegraphics[width=7cm]{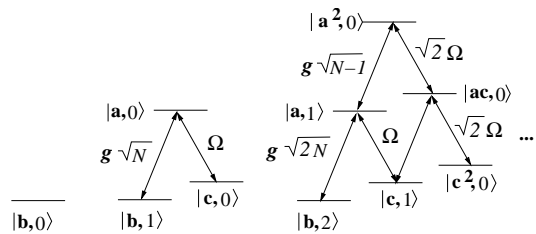}
\caption{
Coupling of
bare eigenstates of atom plus cavity system for at most two photons.
}
\label{3-level-collective}
\end{figure}

When all atoms are initially prepared  in  level $|b\rangle$ 
the only states coupled by the interaction are the totally symmetric
Dicke-states \cite{Dicke54} (after absorption of the spatial phase
factors into the definition of the states)
\be
|{\bf b}\rangle_N &=& |b_1,b_2\dots b_N\rangle,\label{b}\\
|{\bf a}^1\rangle_N &=& \frac{1}{\sqrt{N}} \sum_{j=1}^N
|b_1\dots a_j\dots b_N\rangle,\label{a1}\\
|{\bf c}^1\rangle_N &=& \frac{1}{\sqrt{N}} \sum_{j=1}^N
|b_1\dots c_j\dots b_N\rangle,\label{c1}\\
|{\bf c}^2\rangle_N &=& {N \choose 2}^{-1/2}
\sum_{i< j=1}^N
|b_1..c_i..c_j..b_N\rangle,\label{c2}\\
&&\quad{\rm etc.} .\nonumber
\ee
The couplings within the sub-systems corresponding to 
a single and a double excitation 
are shown in  Fig.~\ref{3-level-collective}.
The set of collective states can be separated into groups with 
a specific excitation number $n$ and atom number $N$. Due to the 
symmetry of the interaction there is no coupling between classes
with different excitation. Decay out of the excited state 
and therefore out of the relevant subsystem couples
only classes with different atom number.


In the following we will restrict ourselves to two-photon resonance, i.e.
$\omega = \omega_a -\omega_c -\nu$. Furthermore for simplicity single-photon
resonance is assumed as well. 
This is sufficient since here we are not
interested in the fidelity of the transfer process itself.
The influence of a finite two-photon detuning on the transfer process
is discussed in detail in \cite{Mewes-PRA-2002}.
In the case of two-photon resonance, the interaction of the 
$N$-atom system with the quantized
radiation mode has a family of dark-states, i.e. adiabatic eigenstates with
vansihing component of the excited states $|a\rangle_j$
\be
&&|D,n\rangle_N = 
\sum_{k=0}^n\xi_{nk}(-\sin\theta)^k (\cos\theta)^{n-k}
|{\bf c}^k,n-k\rangle_N,\nn\\
&&\qquad\xi_{nk}\equiv 
\sqrt{\frac{n!}{k!(n-k)!}},
 \quad\tan\theta(t) \equiv\frac{g\sqrt{N}}{\Omega(t)}.\label{darkn}
\ee
It should  
be noted that 
although the dark states $|D,n\rangle_N$ are degenerate,
there is no transition between them even if 
non-adiabatic corrections are taken into account
due to the symmetry of the interaction Hamiltonian.
Adiabatically rotating the mixing angle $\theta$ from $0$ to 
$\pi/2$ leads to a complete and reversible transfer of all 
photonic states to a collective atomic
excitation if the maximum number of photons $n$ is less than the number
of atoms  $N$. 
If the initial quantum state of the single-mode light field 
is described by the density matrix 
$\hat\rho_f=\sum_{n,m} \rho_{nm}\, |n\rangle\langle m|$, the 
transfer process generates a quantum state of collective excitations
according to
\be
&\sum_{n,m}\rho_{nm} \, |n\rangle\langle m|
 \otimes |{\bf b}\rangle_{N\, N}\langle{\bf b}| &\nonumber \\
&\downarrow &\nn \\
& \sum_{n,m}\rho_{nm} \,|D,n\rangle_{N\, N}\langle D,m|&\\
&\downarrow &\nn \\
&|0\rangle\langle 0| \otimes  
\sum_{n,m} \rho_{nm}\, |{\bf c}^n\rangle_{N\, N}\langle {\bf c}^m|&.
\nonumber
\ee
The dark states of the $N$-atom system 
can be identified as quasi-particle excitations of the so-called
dark-state polaritons $\Psi$  in the
space of atoms and cavity mode \cite{Fleischhauer-PRL-2000}
\be
|D,n\rangle_N = \frac{1}{\sqrt{n!}} \Bigl(\Psi^\dagger\Bigr)^n
|{\bf b},0\rangle_N,
\ee
where $|{\bf b}\rangle$ is the total ground state of the $N$ atom
system and $|0\rangle$ the vacuum state of the cavity mode.
The dark-state polariton  defined as
\be
\Psi = \cos\theta(t)\, a -\sin\theta(t)\, \frac{1}{\sqrt{N}}
\sum_{j=1}^N \sigma_{bc}^j,\label{Psi-def}
\ee
is a superposition of the resonator mode and the collective spin 
corresponding to the ground-state transition $|b\rangle
\leftrightarrow|c\rangle$. 
Associated with the dark polariton
is a bright polariton 
\be
 \Phi_0 = \sin\theta(t)\, a +\cos\theta(t)\, \frac{1}{\sqrt{N}}
\sum_{j=1}^N \sigma_{bc}^j.\label{Phi0-def}
\ee
To obtain a complete set of operators in the space of the cavity mode and
the $N$ atoms in internal states $|b\rangle$ and $|c\rangle$
we also need to introduce the operators
\be
\Phi_l =
\frac{1}{\sqrt{N}}
\sum_{j=1}^{N} \sigma_{bc}^j \, \exp\left\{2 \pi i \frac{l j}{N}\right\}
,\quad l=1 \dots N-1,\label{Phi-def}
\ee
together with the hermitian adjoints $\Psi^\dagger$ and $\Phi_l^\dagger$.
We will also refer to the $\Phi_l$'s as bright 
polaritons. 
In the limit of small atomic excitations
 the polariton operators
obey approximately bosonic commutation relations
\be
[\Psi,\Psi^\dagger]  &=& \cos^2\theta + \sin^2\theta 
\frac{1}{N}\sum_{j=1}^N\Bigl(\sigma_{bb}^j
-\sigma_{cc}^j\Bigr)  \nn \\
&=&   1 +{\cal O}\left(\frac{n_c}{N}\right), \label{commutator1}
\ee
\be
[\Phi_0, \Phi_0^\dagger] 
 &=&  \sin^2\theta + \cos^2\theta 
 \frac{1}{N}\sum_{j=1}^N\Bigl(\sigma_{bb}^j
 -\sigma_{cc}^j\Bigr) \nn \\
&=& 1 +{\cal O}\left(\frac{n_c}{N}\right), \label{commutator2}
\ee
\be
[\Phi_l , {\Phi}^\dagger_m ]  &=& \delta_{lm}
\frac{1}{N}\sum_{j=1}^N\Bigl(\sigma_{bb}^j
-\sigma_{cc}^j\Bigr) \nn\\
&=&  \delta_{lm}  +{\cal O}\left(\frac{n_c}{N}\right),
\label{commutator3}
\ee
where $n_c =\langle \sum_j\sigma_{cc}^j\rangle \ll N$
is the total population in level $|c\rangle$.
Polariton operators of different type commute in lowest order of
$n_c/N$:
\be
[\Phi_i,\Psi^{(\dagger)}]= {\cal O}\left(\frac{n_c}{N}\right).
\label{commutator4}
\ee
It should be noted that the dark and bright polariton 
operators are explicitely time dependent throught the mixing
angle $\theta(t)$.

The collective storage 
state corresponding to a coherent state of light factorizes 
as can be seen quite easily
\begin{eqnarray}
&{\rm e}^{|\alpha|^2/2}\, |\mathbf{b},\alpha\rangle_N =
{\rm e}^{\alpha\,  a^\dagger} |\mathbf{b},0\rangle_N &\nonumber\\
&\downarrow &\nonumber\\ 
&{\rm e}^{\alpha\,  \Psi^\dagger} |\mathbf{b},0\rangle_N&\\
&\downarrow &\nonumber\\ 
&\exp\left\{-\frac{\alpha}{\sqrt{N}}\, \sum_j \sigma_{cb}^j\right\}  
|\mathbf{b},0\rangle_N =  \prod_j \left(1-\frac{\alpha\sigma_{cb}^j}{\sqrt{N}} 
 \right) \,  |\mathbf{b},0\rangle_N.&
\nonumber
\end{eqnarray}
On the other hand, storage states corresponding to non-classical states
of light such as Fock states are maximally entangled $N$ particle 
states $|\mathbf{c}^n\rangle_N$ as can be seen 
from eq.(\ref{c1}),(\ref{c2}). These states 
are known to be rather sensitive to decoherence
processes. For example if for an initial state $|{\bf c}^1\rangle_N$ 
the atom number one undergoes a 
transition from level $|b\rangle$ to an auxiliary state, say  
$|d\rangle$, the resulting state is almost orthogonal to the original
one
\be
&
|c_1,b_2\dots b_N\rangle +|b_1,c_2\dots b_N\rangle
+\cdots |b_1,b_2\dots c_N\rangle &\nonumber\\
&\downarrow &\\
&
|c_1,b_2\dots b_N\rangle +|d_1,c_2\dots b_N\rangle
+\cdots |d_1,b_2\dots c_N\rangle &\nonumber.
\ee
If $p$ denotes the probability of one atom to undergo
a transition from $|b\rangle$ to $|d\rangle$ due to environmental
interactions, the total probability $P_{\rm error}$ to end up in
an orthogonal state scales as $P_{\rm error}\sim 1- (1-p)^N \sim p N$. 
Thus one might naively expect that for the storage of a single photon
the collective quantum memory
will have an $N$ times enhanced sensitivity to decoherence as compared
to a single-atom device. We will now show that this conclusion is 
generally not correct. 

From the inverse relation
\be
a = \cos\theta(t)\, \Psi +\sin\theta(t)\, \Phi_0
\ee
one recognizes that for the resonator mode only excitations
of the dark polariton $\Psi$ and the bright polariton $\Phi_0$ 
matter. Furthermore if after the storage of photon states in
the atomic system the electromagnetic excitations are regenerated
by rotating $\theta$ back from $\pi/2$ to $0$, only
excitations in the dark polariton mode are relevant. I.e. if $W$ denotes
the total density operator of the combined atom-cavity system
after the writing process, only the reduced density operator
\be
\rho= {\rm Tr}_{\Phi}\Bigl\{ W\Bigr\}\label{reduced}
\ee
is relevant for the storage. Here $\textrm{Tr}_\Phi$ denotes the
partial trace over all bright-polariton excitations.
For this reason all states of the 
total system that have the same number of dark-polariton
excitations but an arbitrary number of excitations in
any bright-polariton mode are {\it equivalent} from the point
of view of storage. I.e. there exist equivalence classes
of storage states of the form
\be
|D,n\rangle_N \, \triangleq
\, \biggl\{ \left(\Phi_i^\dagger\right)^k \left(\Phi_j^\dagger\right)^l
\dots \left(\Psi^\dagger\right)^n\, |{\bf b},0\rangle_N\biggr\}.
\ee
It is important to note that any perturbation that acts only onto
bright polariton modes does also not destroy 
superpositions of storage states, since $[\Phi_l,\Psi]=0$:
\be
&&\Phi^\dagger_l\, \sum_n\alpha_n |D,n\rangle_N 
=\Phi^\dagger_l\, \sum_n\frac{\alpha_n}{\sqrt{n!}} 
\left(\Psi^\dagger\right)^n
|{\bf b},0\rangle_N \nonumber\\
&&\qquad\qquad=\sum_n\frac{\alpha_n}{\sqrt{n!}} \left(\Psi^\dagger\right)^n\, 
 \left\{\Phi^\dagger_l\, |{\bf b},0\rangle_N\right\}.
\ee
Likewise all dark
states with the same number of excitations $n$ but with different 
number of atoms $N\gg n$ are equivalent because in the adiabatic read-out
process all dark-polariton operators corresponding
to different $N$ have the same asymptotic mapping $\Psi\to a$
for $\theta\to 0$.
This will be important later on 
when discussing the effect of atom losses from the system
\be
|D,n\rangle_N \, \triangleq \, |D,n\rangle_{N^\prime},\qquad{\rm if}\quad
N,N^\prime \gg n.
\ee
The importance of the equivalence classes stems from the fact
that unwanted interactions with the environment which lead only to 
transitions within the equivalence classes and that do not destroy the
relative phase between them do not affect the
fidelity of the quantum memory.

We can now express the complex Hamiltonian (\ref{ham}) in terms of bright
and dark polariton operators after adiabatically eliminating the
excited state $|a\rangle$.
Separating the oscillatory factor ${\rm e}^{-i\nu t}$
by a canonical transformation and assuming two
photon resonance, i.e. $\omega=\omega_c+\nu$
we arrive at
\be
H&=&\hbar\omega \Bigl(
\Psi^\dagger \Psi+\sum_{l=0}^{N-1}\Phi_l^\dagger\Phi_l\Bigr)
-i\hbar \frac{\Omega^2(t)}{\gamma} \sum_{l=1}^{N-1} \Phi_l^\dagger\Phi_l
\label{ham-pol}\\
&=&\hbar\omega \Bigl(
\Psi^\dagger \Psi+\sum_{l=0}^{N-1}\Phi_l^\dagger\Phi_l\Bigr)
-i\hbar \frac{g^2 N}{\gamma} \cot^2\theta(t)
\sum_{l=1}^{N-1} \Phi_l^\dagger\Phi_l.\nonumber
\ee
We here see two important points. First of all the
adiabatic dynamics does not couple different polariton modes.
Secondly all bright polariton excitations $\Phi_l$, $l=1,2,\dots,N-1$
decay by optical pumping, i.e. by excitation to the excited
state and successive spontaneous emission if $\theta\ne \pi/2$, 
while the dark polaritons
$\Psi$ as well as the bright polaritons $\Phi_0$ are immune to spontaneous
emission. 
Since the non-hermitian Hamiltonian is expressed in terms
of explicitly time dependent variables it describes the dynamics
only in the adiabatic limit \cite{Fleischhauer-PRA-2002}
\be
g\sqrt{N}\, T \gg 1,
\ee
where $T$ is a characteristic time of changes. If non-adiabatic
corrections are taken into account, dark and bright polariton 
modes are coupled with a rate proportional to $\dot\theta$.


\section{Influence of Decoherence on storage fidelity}



\subsection{Imperfect preparation}


In the description of the storage process given in the last section we 
had assumed that every atom in the ensemble was prepared in the ground state 
$|b\rangle$. If the ensemble is large enough the propbability to find an
atom e.g. in state $|c\rangle$ will however be non-negligible, e.g. due to
the interaction with a finite temperature reservoir. Naively one 
might expect that any atom left in state $|c\rangle$ after preparation of the 
ensemble would mimic a stored photon. This would require to make the 
initial probability to find an atom in state $|c\rangle$ small compared
to $1/N$, which is however not the
case. It is rather sufficient that the initial probability of excitation of
a dark-state polariton is small compared to unity. If we consider e.g.
as initial state a thermal state of temperature $\beta=1/k_B T$
$(\theta=\pi/2)$
\begin{eqnarray}
\rho_0=
\frac{1}{Z}\exp\left\{-\beta \hbar\omega_c \left(\hat\Psi^\dagger
\hat\Psi +\sum_{l=1}^{N-1}\hat \Phi_l^\dagger \Phi_l\right)\right\}
\end{eqnarray}
with $Z$ being the statistical sum,
the mean number of initially excited dark-state polaritons
is independent on $N$ and given by
\begin{eqnarray}
\langle \hat\Psi^\dagger\hat \Psi\rangle = 
\frac{{\rm e}^{-\beta\hbar\omega_c}}{1-
{\rm e}^{-\beta\hbar\omega_c}}=\frac{1}{N}
\sum_{i=1}^N \left\langle \sigma_{cc}^j\right\rangle
\end{eqnarray}
Thus if the probability that an atom is initially 
in level $|c\rangle$ is small 
compared to unity, the number of initial dark-polariton excitations
is small compared to unity as well.


\subsection{Random spin flips and dephasing}


On the level on individual atoms the storage occurs within the
two-state system consisting of $|b\rangle$ and $|c\rangle$.
If we assume that all other atomic states including $|a\rangle$ are
energetically much higher, we may safely neglect decoherence 
processes involving the excitation of those states. Then decoherence
caused by individual and independent reservoir interactions 
can be described by the action of the two-level Pauli operators
\begin{eqnarray}
 X_j= \sigma_{bc}^j+\sigma_{cb}^j,\enspace  Z_j = \left[\sigma_{bc}^j,
\sigma_{cb}^j\right],\enspace  Y_j= i \sigma_{bc}^j-i\sigma_{cb}^j.\, 
\end{eqnarray}
$X_j$ describes a symmetric spin flip of the $j$th atom, 
$Z_j$ a phase flip, and $Y_j$ a 
combination of both. Any single-atom error can be 
expressed in terms of these and we will restrict the discussion 
here to the action of $X_j$ (symmetric spin flip), $X_j+i Y_j$ 
(asymmetric spin flip) and $Z_j$ (phase flip).

Inverting relations (\ref{Psi-def}), (\ref{Phi0-def}), and (\ref{Phi-def})
one easily finds a representation of $\sigma_{cb}^j$ in terms of polaritons
\be
\sigma_{cb}^j 
&=&\frac{1}{\sqrt{N}} \left(
\sum_{l=1}^{N-1} \exp\left\{-2\pi i\frac{lj}{N}\right\}
\, \Phi_l^\dagger -\Psi^\dagger\right)\nn\\
&=&\frac{1}{\sqrt{N}} \left(
\sum_{l=1}^{N-1} \eta_{jl}
\, \Phi_l^\dagger -\Psi^\dagger\right)= \frac{X_j+ i Y_j}{2}.\label{sigma}
\ee
Here and in the following 
$\theta=\pi/2$ is assumed, unless stated otherwise, 
which corresponds to the case
of a completed transfer from the radiation field to the ensemble.
Furthermore
\be
X_j=\frac{1}{\sqrt{N}}\left[\sum_{l=1}^{N-1}\left(
\eta_{jl}\Phi_l^\dagger + \eta_{jl}^* \Phi_l\right) - \Psi - \Psi^\dagger
\right].\label{X}
\ee
and
\be
Z_j = \frac{1}{N}\left[\sum_{l=1}^{N-1}\eta_{jl}^*\Phi_l-\Psi,
\sum_{m=1}^{N-1}\eta_{jm}\Phi_m^\dagger-\Psi^\dagger\right].
\label{Z}
\ee
One recognizes at this point that applying the approximate commutation
relations (\ref{commutator1}-\ref{commutator4}), 
which have been obtained with the assumption
$\sigma_{bb}^j \approx 1$ and $\sigma_{cc}^j\approx 0$, would here lead to 
$Z_j=\mathbf{1}_j$. Thus care must be taken when using $Z_j$.


\subsubsection{Spin flip from $|b\rangle \to |c\rangle$}


Consider a quantum memory initially in an ideal storage state $W_0$, 
i.e. without bright polariton excitations. Suppose 
an  atom then
undergoes a spin flip to the internal state $|c\rangle$ if it is 
initially in state $|b\rangle$. 
Such a spin flip process, which could mimic a stored photon, can be 
described by the positive map
\be
W_0 \rightarrow W_1 = \frac{\sigma_{cb}^j W_0 \sigma_{bc}^j}
{{\rm Tr}\{\sigma_{cb}^j W_0 \sigma_{bc}^j\}}.\label{W-map}
\ee
As noted in the previous section only the reduced
density operator traced over the bright-polariton modes is of relevance
for the storage process. Carrying out this trace yields
\be
&&\textrm{Tr}_\Phi\Bigl(\sigma_{cb}^j W_0 \sigma_{bc}^j\Bigr)
= \frac{1}{N} \textrm{Tr}_\Phi\Biggl[
\sum_{l,m}^{N-1} \eta_{jl}\Phi_l^\dagger
W_0 \eta_{jm}^*\Phi_m
\Biggr]\label{map-sigma} \\
&&\qquad
+\frac{1}{N} \Psi^\dagger \rho_0 \Psi-\frac{1}{N}\textrm{Tr}_\Phi\left[
\sum_l^{N-1} \eta_{jl}\Phi_l^\dagger  W_0
\Psi + h.a.\right].\nn
\ee
where $\rho_0=\textrm{Tr}_\phi\{W_0\}$. If we make use of the fact that
the bright and dark polaritons commute in first order of $1/N$ we see that 
the last term in eq.(\ref{map-sigma})
vanishes since there are no excitations of bright polaritons
in the initial state $W_0$. For the same reason 
\be
\textrm{Tr}_\Phi\{\eta_{jl}\Phi_l^\dagger W_0 
\eta_{jm}^*\Phi_m\} = \rho_0\, \delta_{lm},
\ee
and the first term in (\ref{map-sigma}) evaluates to $(1-1/N)\rho_0$. 
Thus we arrive at
\be
\textrm{Tr}_\Phi\Bigl(\sigma_{cb}^j W_0 \sigma_{bc}^j\Bigr) 
= \Bigl(1-\frac{1}{N}\Bigr)\rho_0+\frac{1}{N} \Psi^\dagger \rho_0 
\Psi, \nonumber
\ee
and 
\be
\rho_1= \textrm{Tr}_\Phi\bigl\{W_1\bigr\}= 
\frac{\left(1-\frac{1}{N}\right)\rho_0 +\frac{1}{N}
\Psi^\dagger \rho_0 \Psi}
{1 + \frac{1}{N}
\left\langle\Psi^\dagger \Psi\right\rangle}.\label{rho-1}
\ee
One recognizes that the spin flip of an individual atom only causes
an error of order $1/N$. This exactly compensates for the fact that
the total spin-flip probability of the $N$ atoms 
is $N$ times the probabiliy of a single atom.

From eq.(\ref{rho-1}) one can easily calculate the fidelity
of the quantum memory after a single spin flip error, 
which for the case of an initial pure state 
$\rho_0=|\psi_0\rangle\langle \psi_0|$ is defined as
\be
f\Bigl(|\psi_0\rangle\Bigr)
=\langle\psi_0|\rho_1|\psi_0\rangle = \textrm{Tr}\{\rho_1\rho_0\}.
\ee
One finds
e.g. for a stored Fock-state $|n\rangle$ with $n\ll N$
\be
f_{b\to c}\Bigl(|n\rangle\Bigr) = \frac{1- \frac{1}{N}}{1-\frac{n}{N}} = 1 - \frac{n+1}{N} +{\cal O}\left(\frac{1}{N^2}\right),
\ee
while for a coherent state $|\alpha\rangle$ holds
\be
f_{b\to c}\Bigl(|\alpha\rangle\Bigr) = 
\frac{1-\frac{1}{N} + \frac{|\alpha|^2}{N}}
{1+\frac{|\alpha|^2}{N}} =
 1-\frac{1}{N} +{\cal O}\left(\frac{1}{N^2}\right).
\ee
This reflects the general property of non-classical states to be
more sensitive to decoherence than classical ones.

An alternative way of demonstrating that spin-flip errors do not
depend on the number of atoms is to consider the
Liouville operator ${\cal L}$ describing uncorrelated 
spin flips with rate $\Gamma$
\be 
\dot W &=&{\cal L} W  = \sum_{j=1}^N {\cal L}_j W,\\
{\cal L}_j W &=& -\frac{\Gamma}{2}\left(\sigma_{bc}^j \sigma_{cb}^j 
W + W \sigma_{bc}^j \sigma_{cb}^j -2 \sigma_{cb}^j W \sigma_{bc}^j\right).
\ee
Substituting expression (\ref{sigma}) yields after tracing over the
bright polariton excitations
\be
{\cal L} \rho = - \frac{\Gamma}{2}\left(\Psi^\dagger\Psi \rho +
\rho \Psi^\dagger\Psi -2 \Psi^\dagger \rho\Psi\right).
\ee
One recognizes that the decoherence rate of the reduced density operator
of the ensemble of atoms due to spin flips is the same as for a single atom.


\subsubsection{Symmetric spin flip}


If instead of the asymmetric spin flip $\sigma_{cb}^j = X_j +i Y_j$ 
a symmetric flip happens, eq.(\ref{W-map}) attains the form
\be
W_0 \rightarrow W_1 = \frac{X_j W_0 X_j}
{{\rm Tr}\{X_j W_0 X_j\}}.\label{X-map}
\ee
We here have kept the normalization denominator although ${\rm Tr}
\{X_j W_0 X_j\}=1$ because we want to make use of the approximate
commutation relations between dark and bright polaritons which 
hold only to first order in $1/N$. Thus both, the numerator and denominator
in (\ref{X-map}) have to be expanded in the same way to keep the 
normalization.
Carrying out the trace over the bright polaritons yields
\be
&&\textrm{Tr}_\Phi\Bigl(X_j W_0 X_j\Bigr)\nn\\
&& = \frac{1}{N} \textrm{Tr}_\Phi\Biggl[
\sum_{l,m}^{N-1} \left(\eta_{jl}\Phi_l^\dagger+h.a.\right)
W_0\left(\eta_{jm}\Phi_m^\dagger+h.a.
\right)\Biggr]\nn\\
&&+\frac{1}{N} (\Psi^\dagger+\Psi) \rho_0 (\Psi^\dagger+\Psi)
\label{map-x}\\
&&-\frac{1}{N}\textrm{Tr}_\Phi\left[
\sum_l^{N-1} \left(\eta_{jl}\Phi_l^\dagger+\eta_{jl}^*\Phi_l\right) W_0
(\Psi+\Psi^\dagger) + h.a.\right].\nn
\ee
Again
the last term in eq.(\ref{map-x})
vanishes since there are no excitations of bright polaritons
in the initial state $W_0$ and in the first term 
only the combination 
\be
\textrm{Tr}_\Phi\{\eta_{jl}\Phi_l^\dagger W_0 
\eta_{jm}^*\Phi_m\} = \rho_0 \delta_{lm}
\ee
remains and this term evaluates to $(1-1/N)\rho_0$. This yields
\be
\textrm{Tr}_\Phi\Bigl(X_j W_0 X_j\Bigr) &=& \Bigl(1-\frac{1}{N}\Bigr)\rho_0\\
&&+\frac{1}{N}\left(\Psi^\dagger +\Psi\right) \rho_0\left(\Psi^\dagger 
+\Psi\right),\nonumber
\ee
and we arrive at
\be
\rho_1 = \frac{\left(1-\frac{1}{N}\right)\rho_0 +\frac{1}{N}
(\Psi^\dagger +\Psi)\rho_0 (\Psi^\dagger+\Psi)}
{1-\frac{1}{N} + \frac{1}{N}
\left\langle \left(\Psi^\dagger +\Psi\right)^2\right\rangle},
\ee
which is similar to the case of an asymmetric spin-flip, 
eq.(\ref{rho-1}). Once again it is seen that the
collective quantum memory does not have an enhanced sensitivity
to spin flip errors as compared to a single-atom system.

The fidelity of the memory now reads for a stored Fock and coherent 
state 
\be
f_{b\leftrightarrow c}(|n\rangle)  &=& 1 - \frac{2n+1}{N} 
+{\cal O}\left(\frac{1}{N^2}\right)\\
f_{b\leftrightarrow c}(|\alpha\rangle)  &=& 1 - \frac{1}{N} 
+{\cal O}\left(\frac{1}{N^2}\right)
\ee
%
%


\subsubsection{Phase flip}\label{phase-flip}


If after the preparation of an ideal storage state an atom undergoes
a phase flip the corresponding positive map would read
\be
W_0 \rightarrow W_1=\frac{Z_j W_0 Z_j}{\textrm{Tr}\left\{Z_j W_0 Z_j\right\}}.
\ee
This map is  however not a good starting point of further discussions
because the approximations used when introducing bosonic
polariton operators lead to $Z_j\equiv \mathbf{1}_j$. For this reason we
follow a different approach and calculate the fidelity of the quantum
memory directly. Consider an ideal storage state initially of the form
\be
|\psi_0\rangle =\sum_{n=0}^M c_n |D,n\rangle =
\sum_{n=0}^M
\frac{1}{\sqrt{n!}} 
\bigl(\Psi^\dagger\bigr)^n |\mathbf{b},0\rangle
\label{psi-0-phase}
\ee
where $M\ll N$. 
If the $k$th atom undergoes a phase flip the state changes according to
\be
|\psi_0\rangle \rightarrow |\psi_1\rangle
= \sum_{n=0}^M c_n \frac{1}{\sqrt{n!}} \bigl(\widetilde{\Psi}^\dagger\bigr)^n 
|\mathbf{b},0\rangle
\ee
where
\be
\widetilde{\Psi} \equiv  -\frac{1}{\sqrt{N}}\left[
\sum_{j\ne k} \sigma_{bc}^j - \sigma_{bc}^k\right] = \Psi + \frac{2}{\sqrt{N}}
\sigma_{bc}^k.
\ee
Using eq.(\ref{sigma}) this can be written in the form
\be
\widetilde{\Psi} =
 \left(1-\frac{2}{N}\right)\Psi + 
\frac{2}{N}\sum_{l=1}^{N-1}\eta_{kl}^* \Phi_l.
\ee
This yields in lowest order of $1/N$
\be
\widetilde\Psi^n =\left(1-\frac{2 n}{N}\right) \Psi^n
+\frac{2 n}{N}\sum_l \eta^*_{kl} \Phi_l \Psi^{n-1}
+{\cal O}\left(\frac{1}{N^2}\right).
\ee
Tracing over the bright polariton excitations leads to the
reduced density operator
\be
&&\rho_1 = \textrm{Tr}_\Phi\Bigl[|\psi_1\rangle\langle \psi_1|\Bigr] \\
   &&= \sum_{n,m=0}^M c_n^* c_m
\left(1- \frac{2(n+m)}{N} \right) |D,n\rangle\langle D,m|
 \nn
\ee
and eventually to the fidelity
\be
f_{\rm deph}(|\psi_0\rangle) = 1-\frac{2 \langle n\rangle}{N} +{\cal O}
\left(\frac{1}{N^2}\right),
\ee
where $\langle n\rangle=\langle\Psi^\dagger\Psi\rangle$ 
is the average number of dark-state polaritons in
the initial state. 
One recognizes that a phase flip of a single atom leads to a fidelity
reduction which is of the order of $1/N$. The term $1/N$ again compensates
for the fact that in an $N$-atom ensemble the likelyhood that one arbitrary
atom undergoes a phase flip is $N$ times the probability of a phase flip
for a single atom. It is interesting to note that the fidelity only depends
on the average dark-state polariton number. I.e. dephasing affects in lowest
order of $1/N$ classical and nonclassical states in a similar way.


\subsection{One-atom losses}


Another important source of errors in a collective quantum memory
is the loss of an atom from the ensemble. As discussed in section II
all storage states corresponding to the same  dark-state excitations
in ensembles of different atom number are equivalent as long as the
atom number is large compared to the relevant number of stored photons.
We now calculate the fidelity of the quantum memory after loss
of one atom. We consider again an ideal initial storage state
\be
|\psi_0\rangle_N =\sum_{n=0}^M c_n |D,n\rangle_N,
\ee
where the subscript $N$ denotes the total number of atoms in the ensemble
and $M\ll N$.
The loss of an atom, which, without loss
of generality, can be taken to be the $N$th atom, can be described
by the partial trace over the degrees of freedom of that atom
\be
W_1 = \textrm{Tr}_N\left\{|\psi_0\rangle\langle \psi_0|\right\}.
\ee
To carry out the trace let us first consider the case of a Fock state
of $n$ polaritons $|D,n\rangle_N$
\be
|D,n\rangle_N ={N\choose n}^{-1/2}
\!\!\!\sum_{j_1<\dots < j_n}^N \!\!\bigl| b_1\dots c_{j1}\dots c_{jn}\dots b_N\bigr\rangle.
\ee
Tracing over the $N$th atom results into
\be
&&\textrm{Tr}_N\Bigl\{|D,n\rangle_{NN}\langle D,n|\Bigr\} \nn\\
&& =  {N\choose n}^{-1/2}\!\!\!\sum_{j_1<\dots< j_n}^{N-1}
\bigl|b_1...c_{j1}...c_{jn}...b_{N-1}\bigr\rangle \bigl\langle \dots 
\bigr| \\
&& + 
{N\choose n}^{-1/2}\!\!\! \sum_{j_1<\dots <j_{n-1}}^{N-1}
\bigl|b_1...c_{j1}...c_{j(n-1)}...b_{N-1}\bigr\rangle \bigl
\langle \dots\bigr|
\nn\\
&&= \frac{N-n}{N} \bigl|D,n\bigr\rangle_{N-1 N-1}\bigl\langle D,n\bigr|\nn\\
&&\enspace
+\frac{n}{N} \bigl|D,n-1\bigr\rangle_{N-1 N-1}\bigl\langle D,n-1\bigr|.
\ee
Thus the fidelity of the quantum memory for a Fock state $|n\rangle$ after
loss of a single atom is given by
\be
f_{\rm loss}(|n\rangle) = 1-\frac{n}{N}.
\ee
The decrease of the fidelity again scales only
as $1/N$. This result could of course have been expected as the $n$ 
excitations are equally distributed over all atoms. Thus removing one
reduces the stored information only by the amount $n/N$. 
Generalizing the above result to nondiagonal elements
leads after some calculation to
\be
&&\textrm{Tr}_N\Bigl\{|D,n\rangle_{NN}\langle D,m|\Bigr\} \nn\\
&& =  \frac{\sqrt{(N-n)(N-m)}}{N}  \bigl|D,n\bigr\rangle_{N-1 N-1}
\bigl\langle D,m\bigr|\nn\\
&&\enspace
+\frac{\sqrt{n m}}{N} \bigl|D,n-1\bigr\rangle_{N-1 N-1}
\bigl\langle D,m-1\bigr|.
\ee
Thus the fidelity after the loss of an atom reads for the case of a general
state:
\be
f_{\rm loss}(|\psi_0\rangle) = 1-\frac{1}{N}
\Bigl(\langle\Psi^\dagger\Psi\rangle -\langle\Psi^\dagger\rangle
\langle\Psi\rangle\Bigr)
+{\cal O}\left(\frac{1}{N^2}\right).
\label{loss-2}
\ee
If the initial storage state corresponds e.g. to a coherent state,
the second and third term in (\ref{loss-2}) compensate each other and
the fidelity differs from unity only in order $1/N^2$. Here again the
robustness of classical states becomes apparent.


\subsection{Atomic motion}\label{motion}


Until now it has been assumed that the atoms used in the quantum memory
are at a fixed position during the entire storage time. Since the coupling
of the atoms to the quantum as well as control fields contains however
a spatial phase, see eq.(\ref{ham}), atomic motion results in an effective
dephasing and will lead to a reduction of the fidelity. 
Recently Sun et al.
have argued that inhomogeneities of the atom-light interaction strength 
or in the control field together with atomic motion lead to an increase
of the characteristic decoherence rate by a factor $\sqrt{N}$
 \cite{sunquantph0203072v1}. We thus will analyze
the effect of atomic motion
in the following in more detail. To this end we will
follow the approach of subsection \ref{phase-flip} and describe the
motion by the map of an initially ideal storage state $|\psi_0\rangle$,
(\ref{psi-0-phase}),
according to 
\be
|\psi_0\rangle \rightarrow |\psi_1\rangle
= \sum_n c_n \frac{1}{n!} \bigl(\check{\Psi}^\dagger(t)\bigr)^n 
|\mathbf{b},0\rangle,
\ee
where
\be
\check{\Psi}^\dagger(t) = - \frac{1}{\sqrt{N}}\sum_j \sigma_{bc}^j 
\, \exp\left\{-i\Delta \vec k \cdot\vec r_j(t)\right\},\label{check}
\ee
with $\vec r_j(t)$
denoting the position of the $j$th atom at time $t$ and
$\Delta \vec k = \vec k_1 -k_0 \vec e_z$ is the wavevector difference
between control field and quantized mode. It should be noted that
(\ref{check}) is equivalent to a coupling field with inhomogeneous
phase.

To reduce the
effect of motion in an atomic vapor one could either reduce the temperature
or use a buffer gas of sufficient density. 
In the latter case, which has been used in
room temperature gas-cell experiments 
\cite{van-der-Wal-Science-2003,Phillips-PRL-2001},
 the free motion is replaced by a diffusion. 
In the following we will restrict the discussion to this 
important case. We then can assume that the phase
\be
\Delta\phi_j(t)\equiv 
\Delta \vec k \cdot\vec r_j(t)
\ee
follows a Wiener diffusion process \cite{Gardiner}:
\be
\frac{{\rm d}}{{\rm d}t} \Delta\phi_j(t) &=& \mu_j(t),\\
\overline{\, \mu_j(t)\, } &=& 0,\\
\overline{\, \mu_j(t)\mu_k(t^\prime)\, } &=& D\delta_{jk}\delta(t-t^\prime)
\ee
with $D$ being a characteristic diffusion rate. We now want to show that the
decrease in fidelity due to the phase diffusion is only determined
by $D$ and independent of the number of atoms $N$. For this it is sufficient
to consider an initial Fock state $|n=1\rangle$. 
Reexpressing the single-atom flip operators by collective ones yields
\be
\check \Psi^\dagger =\frac{1}{N}\sum_{j} {\rm e}^{i\Delta\phi_j}
\Bigl[-\sum_{l=1}^{N-1}\eta_{jl} \Phi_l^\dagger + \Psi^\dagger\Bigr].
\ee
With this one finds for an inital Fock state $W_0=|D,1\rangle\langle D,1|$
\be
W_1(t)=\overline{\,\Bigl(\check \Psi^\dagger\Bigr)
|\mathbf{b},0\rangle\langle
\mathbf{b},0| \Bigl(\check \Psi\Bigr) \,},
\ee
where the overline denotes averaging over the phase diffusion process. 
From $W_1(t)$ we can calculate the fidelity of the quantum memory 
by first tracing over
the bright polaritons and then sandwiching with the original state 
$|D,1\rangle$. This yields
\be
&& f_{\rm motion}\Bigl(|1\rangle\Bigr) 
=\overline{\, \langle D,1|\, \textrm{Tr}_\Phi\Bigl(W_1(t)\Bigr)\,
|D,1\rangle\, },\nn\\
&&\qquad= \overline{\biggl(\frac{1}{N}\sum_j {\rm e}^{i\Delta\phi_j}\biggr)
\biggl(\frac{1}{N}\sum_k {\rm e}^{-i\Delta\phi_k}\biggr) }\\
&&\qquad = \frac{1}{N} +\frac{1}{N^2}\sum_{j\ne k} \,\,
\overline{\, {\rm e}^{i\Delta\phi_j}\,} \,\,\,
\overline{\, {\rm e}^{-i\Delta\phi_k}\,} \nn\\
&&\qquad = \frac{1}{N}\Bigl(1+(N-1) {\rm e}^{-D t}\Bigr)\sim {\rm e}^{-D t}.\nn
\ee
A generlization to an arbitrary fock state $|D,n\rangle$ leads to 
a fidelity decay proportional to $\exp\{-n D t\}$.
One recognizes that the atomic motion causes a decay of the fidelity
with a rate given only by the single-atom diffusion rate $D$. 
In contrast to the results of Sun, Yi and You \cite{sunquantph0203072v1} 
we find that there is no enhancement of the decay 
with increasing number of atoms, which
is again due to the existence of equivalence classes.


\section{Non-adiabatic coupling and decoherence}


In section \ref{SECTIONII} it was shown that for the retrieval of
a stored quantum state of light  only the
reduced density operator (\ref{reduced}) is relevant. For this
reason all states of the system which have the same number of dark
state polariton excitations but an arbitrary number of excitations
in the bright state polariton modes belong to the same equivalence
class and lead to the same
result, provided the read-out process is adiabatic.
 Due to decoherence a large number of bright state
polaritons may be excited in the system after the
storage period. Now the question arises what
happens to these excitations if the
read-out process is not adiabatic. Even if there is only a weak
non-adiabatic coupling between bright and dark state polariton
modes it may be sufficient to transfer some of the unwanted
excitations into the dark-polariton mode. 
We will show in the following that 
only the bright-polariton
mode $\Phi_0$ can lead to non-adiabatic contributions to
the read-out signal.
For this we consider the (imaginary) interaction 
Hamiltonian eq.(\ref{ham-pol}) in a rotating frame
and add the coupling of the quantized resonator mode $a$ to
free-space modes $ b_k$:
\be
H=-i\hbar \frac{g^2 N}{\gamma} \cot^2\theta(t)
\sum_{l=1}^{N-1} \Phi_l^\dagger\Phi_l
+\hbar \sum_k \kappa a^\dagger b_k + h.a.
\ee
The equation of motion for the dark-state polariton operator
$\Psi=\cos\theta(t) a -\sin\theta(t) \Sigma_{bc}$,
with $\Sigma_{bc}=\sum_j\sigma_{bc}^j/\sqrt{N}$ then reads
\be
\dot \Psi &=& -\dot\theta(t)\, \sin\theta(t)\, a
+\dot\theta(t)\, \Sigma_{bc} + \frac{i}{\hbar} \Bigl[H,\Psi\Bigr] \nn\\
&=& -\dot \theta(t) \, \Phi_0 + i \kappa\cos\theta(t) \sum_k \, b_k.
\ee
Thus the dark-polariton is coupled to the outside modes $b_k$ and the
bright-polariton operator $\Phi_0=\sin\theta(t)  a
+\cos\theta(t) \Sigma_{bc}$ only. In a similar way one finds for
$\Phi_0$ and the $b_k$'s:
\be
\dot\Phi_0 &=& \dot\theta(t) \, \Psi ,\\
\dot b_k &=& i\kappa \, a \nn\\
         &=& i\kappa \cos\theta(t)\, \Psi +i\kappa \sin\theta(t)\, \Phi_0.
\ee
One recognizes that $\{\Psi,\Phi_0,b_k,\}$ are a closed set of coupled
operators even if non-adiabatic corrections in the read-out proportional 
to $\dot\theta(t)$ are taken into account. Thus only decoherence
induced excitations generated in $\Phi_0$ will influence the read-out signal,
all the other $N-1$ bright polariton modes remain uncoupled from the storage
systems.


\section{Summary}


In the present paper we have studied the influence of individual
decoherence processes
on the fidelity of a quantum memory for photons based on ensembles
of atoms. Despite the fact that the atomic storage states corresponding
to non-classical states of the radiation field are entangled many-particle
states, the system shows no enhanced sensitivity to decoherence as compared
e.g. to single-atom storage systems, if it is caused by
a coupling of the atoms to individual and independent reservoirs.
This is due to the existence of equivalence classes of storage states
corresponding to the excitations of only one eigenmode of the system, the
dark-state polariton. It was shown that all states with equal
reduced density operator after tracing out the $N$ bright-polariton
modes reproduce the same quantum state of light in the read-out
process. For similar reasons no stringent requirements for 
preparation of the atomic system before the storage exist.
It is sufficient that the number of atoms
remaining in the storage level $|c\rangle$ after preparation of the
ensemble is small compared to the total number of atoms, which can
easily be achieved by optical pumping. It was  shown moreover that the 
loss of an atom from the sample causes only an error of the order of $1/N$.
Motion of atoms during the storage time causes an effective dephasing
and thus leads to a decrease in fidelity. It was shown, however,
that the corresponding error is independent on the number of atoms, which is
in contrast to the result of \cite{sunquantph0203072v1}. Finally 
since non-adiabatic effects only couple one of the bright polariton modes
to the dark-polariton, the potentially large number of excitations
in the $N$ bright-polariton modes caused by decoherence processes does not
leak into the read-out signal in a significant amount
even if the read-out process is
not adiabatic. 

The present paper proves that atomic ensembles
are suitable systems for the storage of quantum states of the radiation field
even in the presence of non-cooperative decoherence processes. It should
be noted that this conclusion does however not apply to the 
quantum gate between stored photonic qubits based on dipole blockade
proposed in \cite{Lukin-PRL-2001} 
nor to the photon detection scheme suggested in
\cite{Imamoglu-PRL-2002}.


\section*{Acknowledgement}


The authors would like to thank M. D. Lukin and R. Unanyan for stimulating 
discussions. The support of the DFG through the SPP 1078 
``Quanteninformationsverarbeitung'' and the EU network QUACS is gratefully
acknlowedged. 
C. M. also thanks the Studienstiftung des Deutschen Volkes 
for financial support.


\newpage





\end{document}